# The source of the major solar energetic particle events from super active region 11944

David H. Brooks
College of Science, George Mason University, 4400 University Drive, Fairfax, VA 22030, USA.

Stephanie L. Yardley
Mullard Space Science Laboratory, University College London, Holmbury St. Mary, RH5 6NT, UK.

**One sentence summary:** Elemental abundance measurements by the Hinode and Wind spacecrafts reveal the source of major solar energetic particle events.

**Shock waves associated with fast coronal mass ejections (CMEs) accelerate solar energetic particles (SEPs) in the long duration, gradual events that pose hazards to crewed spaceflight and near-Earth technological assets, but the source of the CME shock-accelerated plasma is still debated. Here, we use multi-messenger observations from the Heliophysics System Observatory to identify plasma confined at the footpoints of the hot, core loops of active region 11944 as the source of major gradual SEP events in January 2014. We show that the elemental composition signature detected spectroscopically at the footpoints explains the measurements made by particle counting techniques near Earth. Our results localize the elemental fractionation process to the top of the chromosphere. The plasma confined closest to that region, where the coronal magnetic field strength is high (a few hundred Gauss), develops the SEP composition signature. This source material is continually released from magnetic confinement and accelerated as SEPs following M-, C-, and X-class flares.**

## INTRODUCTION

One of the central goals of heliophysics is to understand the origins of solar activity and predict its impact on the terrestrial space environment. We need to understand and characterize the processes that form and heat the solar atmosphere and accelerate the solar wind into the heliosphere. From a space weather perspective, we must elucidate the mechanisms that drive solar flares, coronal mass ejections (CMEs), and solar energetic particles (SEPs). The recently launched Parker Solar Probe (PSP) (*1*) and Solar Orbiter (*2*) missions will investigate the Sun from a closer vantage point than ever before [within 0.1 and 0.3 astronomical unit (AU)], allowing further insights into these basic processes. Already the new measurements from PSP are fundamentally altering long-held views. The discovery of magnetic field reversals, dubbed "switchbacks" (*3*), raises questions about the extent to which we can connect activity seen at Earth with dynamics in the source regions on the Sun. For an operational space weather predictive capability, we still need to understand what happens in the near-Earth environment. Field reversals that we see close-in where the probes operate are different than what we see at Earth (*4*). The missions of the Heliophysics System Observatory are essential tools to understand the whole Sun-Earth coupled system.

One important tracer of mass flow through the solar atmosphere is the elemental abundance (plasma composition). A separation (fractionation) of ions and neutrals takes place that preferentially enhances elements that are easier to ionize [first ionization potential (FIP) <10 eV] in the corona, solar wind, and SEPs. Elements such as Fe, Ca, and Si are enhanced to varying degrees (usually factors of 2 to 4) compared to elements such as C, N, and O (*5*). This FIP effect is thought to operate in the chromosphere and is also widespread on stars other than the Sun (*6*).

For solar wind and SEP connection studies, plasma composition is central to linking the in situ data to solar sources. Several simple comparisons have been made in the past (*7*, *8*), and more recently, back-mapping and magnetic field extrapolations have been developed (*9*) and tested on composition data (*10*) to support the inner heliospheric missions over the next several years. The plasma composition studies have used measurements from the Hinode (*11*) Extreme-ultraviolet (EUV) Imaging Spectrometer (EIS) (*12*) to track sources of the slow solar wind interacting with the Advanced Composition Explorer (ACE) observatory. To date, however, there have been no EIS studies of the sources of SEPs or any investigation of what specific features in a solar active region (AR) could be sources.

A key question is whether the coronal material that seeds the SEP population during large, gradual events—those with high ion fluxes and the greatest impact on near-Earth space infrastructure—originates from the same source locations as the solar wind and is simply shock-accelerated later by CMEs. In situ measurements of SEP and solar wind abundances suggest otherwise. In particular, the SEP/solar wind–to–photospheric abundance ratio pattern as a function of FIP is different, and the magnitude of the enhancement for some element ratios is also different (*13*). Observations in situ suggest that intermediate FIP elements such as S (the high FIP element with the lowest FIP) behave differently on open and closed magnetic field lines (*13*), making abundance ratios such as Si/S potentially more useful diagnostics than others such as Fe/O. Currently, the most developed model of the FIP effect suggests that the effect is a result of the ponderomotive force, arising from the reflection/ refraction of magnetohydrodynamic (MHD) waves, acting on plasma ions but not neutrals (*6, 14*). In this model, S is fractionated like a high FIP element on closed magnetic fields and more like a low FIP element on open fields. This is because the MHD waves can achieve resonance on closed fields and that restricts fractionation to the top of the chromosphere where H is beginning to ionize. S is not fractionated in these conditions. In contrast, MHD waves on open field do not achieve resonance and cause fractionation throughout the chromosphere, and S can be fractionated lower down where H is neutral (*15*). Essentially, this picture implies that there is more S in the slow solar wind (and co-rotating interaction regions) than in the magnetically closed corona or gradual SEPs.

Here, we report the likely sources of large, gradual SEP events that occurred between 4 and 12 January 2014. We use Hinode/EIS spectroscopic techniques, together with particle counting observations from the EPACT (Energetic Particles: Acceleration, Composition, and Transport) (*16*) instrument on the Wind spacecraft (*17*), to match remote sensing and in situ measurements of the Si/S abundance ratio between the solar corona and events detected at the L1 Lagrange point. We develop a picture of how the source material is produced and evolves the SEP composition signature that is supported by current theoretical models and coronal magnetic field strength measurements.

**RESULTS AND DISCUSSION**

We selected the January 2014 events to simplify the identification of the sources. As we show below, the events are each associated with a solar flare that occurred in AR 11944, so we have a direct link to the host AR without the need to model the magnetic connection. The ACE observatory (*18*) recorded that the solar wind speed fell below 500 km/s on January 5 and did not increase until January 12. The bulk solar wind composition also increased from approximately photospheric values to enhanced factors of 3 to 4 during this period (as measured by Fe/O ratios). These speeds and abundance ratios are typical of the slow solar wind. AR 11944 was the dominant feature on disk during this time and produced several major (M- and X-class) flares. It was a magnetically complex ($\beta\gamma\delta$/F-type) region, and we show a full disk image taken by the Solar Dynamics Observatory (SDO) (*19*) Atmospheric Imaging Assembly (AIA) (*20*) on January 7 in Fig. 1. We also overlay a potential field source surface (PFSS) extrapolation constructed using the Global Oscillation Network Group (GONG) synoptic data. The PFSS approach, of course, has limitations, especially when applied to a complex and flaring AR. Here, we mainly use the extrapolation to show the large-scale magnetic configuration and connectivity of AR 11944, where PFSS models have been relatively successful (*21*). We have verified that the large-scale structure does not change significantly during the time period of our observations, despite smaller-scale disruptions due to flaring. Note that the extrapolation reveals apparent open flux to the solar East side of the trailing negative magnetic polarity (Fig. 2).

We show the solar x-ray flux evolution from 4th to 9th January in Fig. 3. The data are from GOES-15 (Geostationary Operational Environmental Satellites 15) (*22*). The important flares that generated the SEP events we study are the M4.0, C2.6, and X1.2 events that began around 19:00 UT on the 4th, 9:30 UT on the 6th, and 18:00 UT on the 7th, respectively. Within half an hour, a CME was launched with speeds in the range of 960 to 1800 km/s (*23*). The CMEs shock accelerate the ambient coronal plasma that has been continually produced in the AR from, as yet, unidentified sources.

Figure 4 shows that the SEP events were detected by Wind, and we also highlight the times of the associated flares. We see that the particle flux increases almost immediately after the three flares and stays high, in the Si 3.2- to 4.9-MeV channels, for at least a day in all cases. These are classified as gradual events because of the CME association and long (day time scale) duration of high-energy (above tens of megaelectron volts) particles. The fast development of the SEPs at Wind suggests that they were near-Sun CME accelerated shocks (*24*), and magnetic reconnection in the train of consecutive flares (see Fig. 3) is possibly the way that the required population of seed particles is produced that are then injected into the acceleration process (*25*). We see increasing variability in the intensities in Fig. 4, over several orders of magnitude, which could also be an indication of a growing availability of suprathermal particles following each flare (*26*).

Because of the low abundance of S, however, Wind does not detect a measurable signal in the 3.2- to 5.0-MeV channel for the first event, but the signal is strong after the C2.6 flare and has increased by an order of magnitude. It increases by more than two orders of magnitude after the X1.2 flare, so we are able to obtain the evolution of the Si/S abundance ratio for the second and third SEP events (see Materials and Methods). We show this ratio in Fig. 5. Note that the uncertainties in the event averages for the Si/S abundance ratio are very low: <10% for the first and second event and <1% for the third event. The hourly data have higher uncertainties with a mean value of ~25%.

The Si/S SEP abundance ratio shows a range of values but is almost always elevated above photospheric levels throughout both the measured events. The event averages are close to 2.0, which is typical of SEP abundances (*13*), and is at the high end of daily averaged values recorded in the bulk solar wind by ACE during the previous solar cycle 23, which generally fall in the range of 1 to 2 (*27*). Much higher Si/S ratios are seen periodically during these events, reaching values of 4 to 5 at times. We explore where the plasma that has abundances enhanced to these levels originates from using the EIS data.

Hinode/EIS tracked AR 11944 from 4th to 16th January mostly in flare observations mode. Because of telemetry constraints, only a subset of the EIS wavelength range is downloaded when attempting to catch flares, but several more detailed diagnostic scans were also run during the period when the SEP events occurred. The details of these observing sequences and the data reduction and analysis techniques (Doppler velocity and composition maps) are given in Materials and Methods. To provide further context for the composition measurements from these observations, we show several AIA images in Fig. 6. The figure shows the same field of view (FOV) as in Fig. 1, and the images show the structure of the AR at ~0.9, 2.8, and 7.1 MK. Bright "moss" emission (*28*, *29*) is clearly visible in the AR core in the 171-Å image, with high-lying loop arcades connecting to AR 11946 to the North and AR 11943 to the solar West. Fan structures also extend from the East side, where the PFSS extrapolation suggests that the magnetic field is open (Fig. 2). It appears that even the major flares that occurred did not drastically change the large- scale magnetic topology created by the interaction of the three ARs. Note the similarity between the AIA 193-Å image in Fig. 1 and the 171-Å image in Fig. 6. These two images were taken 18 hours before and 17 hours after the X1.2 flare at 18:00 UT on the 7th. The higher temperature AIA 94-Å channel is dominated by Fe XVIII in ARs, but there are also contributions from lower temperature emission lines (*30*). We have isolated the Fe XVIII component using a technique that estimates the amount of lower temperature emission from a combination of simultaneous AIA 171-Å and 193-Å images (*31*). The extracted Fe XVIII image shows that hot loops dominate the emission in the core of the AR above the moss.

Figure 7 shows data from the three EIS scans we used for the composition analysis. An FOV of 240″ by 512″ was scanned using the 1″ slit. The first two scans captured both AR 11944 and 11946 in the same FOV, but the third one was targeted ∼240″ south into the core of AR 11944. We show the FOVs of the EIS scans in Fig. 6. Figure 7 shows intensity and relative Doppler velocity maps obtained from the Fe XII 195.119-Å spectral line formed at the same temperature as the AIA 193-Å images. We have overlaid contours from the Si/S composition maps in red and pink. The contours were constructed from EIS composition maps, and these are noisy. We reduced the noise using a Gaussian filter with a full width at half maximum (FWHM) of 6 pixels and only show contours with lengths larger than 25 pixels. In this way, we isolate coherent, recognizable features with a strong signal. Uncontoured areas in the figure either are too noisy or have too weak signals to make accurate measurements. For example, the strong blue-shifted region on the fifth is positioned above a sunspot. The emission from this region is too weak in the critical S X 264.223-Å line. The contours also correspond to the values in the color bars of Fig. 5. Pink shows where in the AR that EIS measures Si/S abundance ratios that correspond to the SEP event averages. We show a range in the color bars of Fig. 5, rather than single values, to take account of the uncertainty in the EIS composition measurements. These are dominated by the instrument calibration and are ∼30%. Red shows where in the AR that EIS measures Si/S abundance ratios corresponding to the higher values measured in the SEP events (red range in color bars of Fig. 5).

It is clear that the host AR produces plasma with a composition that can explain that measured in situ by Wind. To be precise, the contoured regions in red and pink show Si/S abundance ratios that cover the full range of values detected by Wind.

The red contours mostly surround the bright moss emission in the AR core. Moss corresponds to the footpoints of high temperature (∼4 MK) loops (*32*). It means that plasma with the highest Si/S abundance ratios, a signature that is detected by Wind in the SEP events, is produced at the footpoints of the highest temperature loops in the AR.

High Si/S abundance ratios have been reported at the footpoints of coronal loops before in an anemone AR emerging in a coronal hole (*33*), but these were significantly lower values than we find here (∼2), and the AR did not produce any major SEP events (*34*). Still, it raises the question as to whether this is a common feature of ARs. In addition, this previous work already suggested that higher abundance ratios might be produced closer to the region where the FIP effect operates, i.e., somewhere in the chromosphere. Our understanding of how the FIP effect operates has advanced more recently. As discussed, in the ponderomotive force model, an enhancement of the Si/S abundance ratio can arise if the FIP effect operates at the top of the chromosphere. Our observations favor this scenario.

The closed field of the hottest loops in an AR is anchored in the core where the photospheric magnetic field is strong. The highest photospheric magnetic field strength ever observed (8.2 kG) was measured in a light bridge that formed in the sunspot adjacent to the moss region in AR 11944 on the next solar rotation (*35*, *36*). Typically, the coronal magnetic field strength is lower, but an unusually strong coronal field could explain why the Si/S abundance ratio can reach values as high as 5 in this AR.

The coronal magnetic field strength is, of course, difficult to measure. Very recently, the capability to map the plane-of-sky component of the coronal magnetic field strength outside the solar disk, in stable conditions, using MHD wave density measurements, was demonstrated (*37*). The off-limb field strengths were measured as 1 to 4 G. With an AR positioned on the limb, this technique could be used to investigate the core magnetic field strength, but not during eruptive events, and line-of-sight superposition would make measurements in specific features difficult. Even more recently, a technique to measure the coronal magnetic field strength from the branching ratio of a specific magnetically induced transition (MIT) to a magnetic quadrupole (M2) transition in Fe X was developed (*38*, *39*). In what will likely be a landmark paper, the method was demonstrated to work on EIS data, which opens up access to a complete solar cycle archive of observations (*40*).

Following these demonstrations, we apply the new EIS method here to a specific physical problem and measure the coronal field strength in the hot loop footpoints that we have identified as the likeliest SEP sources. We describe the details of the technique in Materials and Methods. A critical issue is that the method uses the Fe X 174.53- and 175.263-Å spectral lines to initially measure the electron density, and the 175.263-Å line is often weak. With the data at hand—the exposure time of the spectra we used to make the measurement is 30 s—we were only able to analyze the brightest footpoint regions. To further ensure that there was enough signal for an accurate measurement, we averaged the spectra in the black boxed regions shown within the red contours in Fig. 7. These boxes are 90 arc $\sec^2$ (47.3 $Mm^2$). We find values in the range of 245 to 550 G. Currently, this technique has uncertainties on the order of 60 to 70% (*40*), but it is clear that the confining field strength is on the order of hundreds of Gauss.

Our observations also show that other areas in the AR (pink contours in Fig. 7) can supply plasma with lower (but dominant) Si/S abundance enhancements (around the event averages of ~2). These appear to encompass regions where the highest enhancements (red contours in Fig. 7) are more scattered, such as loop footpoints away from the AR core (left of Fig. 7). They converge around the edges of the moss regions at the hot core loop footpoints. This is where the magnetic field is changing and emission is more variable (*41*, *42*) and is also closer to the blue-shifted upflow regions in our velocity maps. It seems that the material that is registered in situ in the SEP events is a combination of plasma that can more directly escape on open or opening field and plasma that requires another agent such as interchange reconnection or a flare/CME shock. The hot loop footpoints develop a high Si/S abundance ratio because the plasma is confined by strong magnetic field close to the elemental fractionation region. This source appears different from that of the slow solar wind only in the sense that it was formed in special conditions within the AR core loops, but other locations that potentially seed the gradual SEP population are producing plasma that is also escaping from around the same loop footpoints, and these intuitively should form part of the solar wind. A critical measurement would be to determine whether abundance ratios can reach the high values reported here in situ in the solar wind on hourly time scales, although this is difficult to do with S because of its low abundance. If that were possible, it would imply no difference between the (slow) solar wind and the coronal plasma that later becomes SEPs.

## MATERIALS AND METHODS

### Instrument data processing and magnetic field model

We used observations from SDO/AIA, SDO/HMI (Helioseismic and Magnetic Imager) (*43*), the GONG Network, GOES, Wind/EPACT, and Hinode/EIS in this article. The 193-Å AIA images and HMI line-of-sight magnetogram are level 1 data products that were obtained from the JSOC (Joint Science Operations Center) online cut-out service at Stanford University. SDO/AIA provides full-disk images of the solar atmosphere in 10 different wavelengths with a pixel size of 0.6″ and a cadence of 12 s. HMI observes the full-disk surface magnetic field at 6173 Å with a pixel size of 0.5″ and a cadence of 45 s. Both have a spatial resolution of around 1″. We used the SunPy software package (*44*) in Python to create a composite map of the two data products.

The GONG magnetogram synoptic map used to construct the PFSS extrapolation was downloaded directly from the National Solar Observatory's (NSO) GONG data archive. GONG is a network of six ground-based telescopes that monitor the Sun 24 hours a day. GONG provides fully calibrated, full-disk, 2.5″ pixel photospheric magnetic field images of the Sun every minute. The synoptic map we used was last updated at 23:54 UT on January 6—close to the time of the AIA image used in Fig. 2. A total of 6512 magnetograms were used to construct the map. These images have been averaged, and the annual periodic modulation in the polar regions has been corrected for. The images are then remapped into longitude and sine (latitude), and the line-of-sight component is converted into flux density by assuming a radial field at the photosphere. The synoptic maps therefore provide the radial component of the photospheric magnetic field that is used by the PFSS model to obtain the extrapolated magnetic field. We used pfsspy (*45*, *46*) to compute the PFSS model of the magnetic field in spherical coordinates using a 10 × 10 footpoint grid to trace the field lines and a source surface radius of $2.5R_\odot$. We subtract the mean radial field to ensure that the solenoidal condition is satisfied. We also used the SunPy software package to overplot the traced magnetic field lines on the AIA image.

The GOES is a series of satellites that carry an x-ray sensor that observes solar flux in two broadband channels: 0.5 to 4 Å and 1 to 8 Å integrated across the full disk. We downloaded the GOES data directly from NASA by using the Federated Internet Data Obtainer (FIDO) search and retrieval tool in Python. The timing, duration, and GOES class of the solar flares were taken from the Heliospheric Event Knowledge catalog.

We obtained the Wind/EPACT Si time series of intensities from the Coordinated Data Analysis Web at the NASA/Goddard Space Flight Center. These are hourly datasets from the Low Energy Matrix Telescope (LEMT) that are provided for several energetic particle ranges. We only used two energy channels in this work: 3.2 to 4.0 MeV per nucleon and 4.0 to 4.9 MeV per nucleon. Because of its lower abundance, hourly measurements for S are more challenging and are not provided online. We obtained the S intensities and also Si/S event averages for the SEPs we studied in this article directly from the EPACT team.

Hinode/EIS is a moderate spatial resolution (~3″) EUV imaging spectrometer that observes in two wavelength bands covering 171 to 212 Å and 245 to 291 Å. Observers can choose from four slit width options (1″, 2″, 40″, and 266″) using a slit exchange mechanism. The best spectral resolution is 22 mÅ. We obtained the EIS data directly from the Data ARchive and Transition System (DARTS) at the JAXA (Japan Aerospace Exploration Agency) Institute of Space and Astronautical Science (ISAS). We used the SolarSoftware IDL procedure eis_prep to reduce the data for analysis. This code takes account of pixels affected by electrical charge, dust, and hits from cosmic rays. It also removes the charge-coupled device (CCD) dark current pedestal.

We used two different EIS observing sequences for our analysis. For the plasma composition measurements, we used scans of AR 11944 with the 1″ slit covering an FOV of 240″ by 512″ in coarse 2″ steps. The exposure time was 60 s, and 25-wavelength windows were telemetered to ground. This sequence contains all the spectral lines we need for the composition analysis (see below) but does not include the critical lines needed for coronal magnetic field measurements. For that analysis, we used a second observing sequence that takes the full spectral range of the CCD. In this case, the 2″ slit was used to cover an FOV of 120″ by 160″, also in coarse 2″ steps. The exposure time was 30 s.

**EIS Doppler velocity measurements**

Temperature variations around the Hinode orbit lead to changes in the thermal environment of the EIS instrument. One consequence is that the spectrum moves across the face of the CCD, and this variation needs corrected before analysis. An artificial neural network model of this orbital variation was developed early in the mission (*47*), and we use this as a first step for our velocity calibration. The EIS grating is also slightly tilted, and there is a spatial offset between the short- and long-wavelength CCDs. These are also corrected by the neural network model, and the final uncertainties in measured velocities (from Gaussian fitting) are about 4.5 km/s.

No absolute wavelength calibration is possible for EIS, so we need to establish a reference wavelength when computing velocities from measured spectral line centroids. We used the strong Fe XII 195.119-Å spectral line for this purpose. Fe XII 195.119 Å is blended with a weak density-sensitive component at 195.179 Å. We used a double Gaussian function to extract the information for the main component. Properly then, the maps we show are of relative Doppler velocities. Ideally, the FOV of the observations would capture a large area of quiet Sun, and we would use this to obtain a reference wavelength. Our observations, however, are pointed at the core of the AR, so this is not possible. Instead, we simply average the fitted Fe XII 195.119-Å line centroids over the upper 160 pixels for the observations on January 5 and 6, and the whole FOV on January 8. Furthermore, there is often a residual orbital variation of the spectral line centroid across the FOV even after the velocity calibration procedure. This is an indication that the standard neural network model is not accurately representing more recent data, so we corrected this residual effect using custom software we have developed (*48*).

Fortunately, the absolute magnitudes of the Doppler velocities are not central to this work. We only use the velocity maps to identify locations of blue-shifted upflows close to the likely SEP sources.

**EIS plasma composition measurements**

EIS Si X 258.375-Å and S X 264.223-Å observations have been used to measure the Si/S abundance ratio for a decade, and the techniques are now well established (*7*). Here, we compute abundance ratio maps (overlaid as contours in Fig. 7) using the method we developed for mapping sources of the slow solar wind (*8*). First, because the Si X 258.375/S X 264.223 intensity ratio is sensitive to electron temperature and density, we measure these plasma properties using other Fe spectral lines available in our datasets. We measure the electron density using the Fe XIII 202.044/203.826 ratio. Then, we use that density to compute contribution functions, under the assumption of ionization equilibrium, for a selection of Fe lines that span the 0.52- to 5.5-MK temperature range. Specifically, we use the following list of Fe lines (with quiet Sun peak formation temperatures shown in parenthesis): Fe VIII 185.213 Å (0.52 MK), Fe VIII 186.601 Å (0.52 MK), Fe IX 188.497 Å (0.87 MK), Fe IX 197.862 Å (0.91 MK), Fe X 184.536 Å (1.15 MK), Fe XI 188.216 Å (1.38 MK), Fe XII 192.394 Å (1.58 MK), Fe XII 195.119 Å (1.58 MK), Fe XIII 202.044 Å (1.74 MK), Fe XIII 203.826 Å (1.74 MK), Fe XIV 264.787 Å (2.00 MK), Fe XIV 270.519 Å

(2.00 MK), Fe XV 284.160 Å (2.19 MK), Fe XVI 262.984 Å (2.75 MK), and Fe XVII 254.870 Å (5.50 MK). For reference, Si X 258.375 Å and S X 264.223 Å are formed at 1.38 and 1.51 MK, respectively.

Next, we compute the temperature distribution [emission measure (EM)] of the plasma by inverting the well-known equation for the spectral line intensities, i.e.

$$I = A \int \phi(T) G(T, n) dT \tag{1}$$

where $G(T, n)$ is the contribution function, dependent on temperature, $T$, and density, $n$; $\phi(T)dT$ is the EM; $A$ is the elemental abundance; and $I$ is the resultant intensity for a specific atomic transition. The contribution functions include all the relevant atomic transition processes needed to establish the population structure of the particular ion (spontaneous radiative decay, electron collisional excitation/deexcitation, and ionization/recombination). For all the atomic calculations, including the ionization equilibrium, we use the CHIANTI database (*49–51*).

The EM distribution is constructed using the Markov-Chain Monte Carlo (MCMC) software in the PINTofALE package (*52, 53*). We perform 100 MCMC simulations to find the EM distribution that best reproduces the observed line intensities. Once the temperature distribution has been found, we adjust it to reproduce the observed Si X 258.375-Å intensity while ensuring that this scaling does not exceed ~40%. Thus, the shape of the temperature distribution is derived from the Fe lines, but the magnitude is determined by the Si line. We use an updated version of the EIS absolute calibration for this work (*54*). Assuming photospheric abundances (*55*), the ratio between the predicted and observed S X 264.223-Å intensity gives us the Si/S abundance ratio. To make the composition maps, we followed this procedure for every pixel in the EIS FOV (9600 to 61440 pixels for the two observing sequences).

Our methodology accounts for the temperature and density sensitivity of the Si/S intensity ratio and any uncertainty in the Fe/Si abundance (or small differences in fractionation levels). There are, however, several other factors that contribute to uncertainties in the measurements. There are generic factors in that we rely on accurate atomic data for all the spectral lines, and we assume that the assumptions we make to calculate the EM are reasonable. In a sense, these uncertainties are coupled: A violation of ionization equilibrium, for example, leads to systematic errors in the $G(T,n)$ functions. We expect that these types of errors would become apparent as systematic differences between observed and calculated intensities. In more practical terms, the EIS instrument calibration has associated uncertainties of 23% (*56*) and has been evolving over time (*54, 57*). Tests show that our method is robust to substantial calibration errors (*8*), but this is the largest source of error in the current study and leads to ~30% accuracy in the EIS composition measurements.

**EIS coronal magnetic field measurements**

The ability to directly measure the coronal magnetic field strength from EIS observations arises from the helpful peculiarities of the $3s^2\ 3p^4\ 3d\ ^4D$ term of Fe X. The fine-structure splitting energy between the $^4D_{7/2}$ and $^4D_{5/2}$ levels along the Cl-like isoelectronic sequence reaches a minimum close to the nuclear charge of Fe. The near degeneracy of these levels allows the production of an MIT from the $^4D_{7/2}$ level to the ground state ($3s^2\ 3p^5\ ^2P^o_{3/2}$) in the presence of an external magnetic field. The theoretical basis of these characteristics is described elsewhere (*38*), and a schematic diagram showing the relevant transitions is also available [Fig. 1 (*39*)].

Because the energy separation between these levels is so small, it is not possible with current spectrometers to resolve the MIT transition from either the M2 magnetic quadrupole forbidden transition to the ground state from the same upper level, or the E1 electric dipole transition to the ground from the $^4D_{5/2}$ level. All of these transitions produce spectral lines close to 257.262 Å. We can, under certain conditions, however, model the spectral line intensity of this feature in a magnetic field–free environment and infer the intensity of the MIT. This can then be compared to the M2 intensity to determine the magnetic field strength because the MIT/M2 intensity ratio has a quadratic dependence on the external field strength. This is the idea behind recent work, and we follow the same method here (*39*, *40*).

First, we inferred the intensity of the MIT transition, $I_{MIT}$, by measuring the excess emission in the Fe X 257.262-Å spectral line. This was achieved by modeling the field-free intensity of the combined E1 and M2 transitions, $I_{E1M2}$, using the Fe X 184.536-Å spectral line. We calculated the theoretical ratio, $r(E1M2/184)$, for the measured electron density using the CHIANTI atomic database and used the observed Fe X 184.536-Å intensity, $I_{184}$, to model the expected value for Fe X 257.262 Å. Subtracting the calculated intensity from the observed one, $I_{257}$, gave the excess emission. The electron density was determined using the Fe X 174.532/175.263 diagnostic ratio.

Second, we inferred the intensity of the M2 transition, $I_{M2}$. This was achieved by calculating the theoretical ratio with Fe X 184.536 Å, $r(M2/184)$, using the measured electron density and using the observed intensity to model the expected value for the M2 component of Fe X 257.262 Å.

The procedure is summarized as

$$I_{MIT}/I_{M2} = \frac{I_{257} - (I_{184} \times r(E1M2/184))}{I_{184} \times r(M2/184)} \qquad (2)$$

Table 1 provides the observed intensities for all the spectral lines we used for these measurements in all of the boxed regions in Fig. 7. We also show the theoretical Fe X 174.532/175.263 ratio dependence on density in Fig. 8A. For the regions we analyzed, densities fall in the range of log ($n/cm^{-3}$) = 9.5 to 9.7. Figure 8B also shows the theoretical $r(E1M2/184)$, $r(M2/184)$, and $r(E1/184)$ ratios as a function of electron density, with the values inferred from the measured densities overlaid.

Last, we compared the $I_{MIT}/I_{M2}$ ratio to the theoretical branching ratio (*38*, *40*) to derive the magnetic field strength. We also show the relationship between the magnetic field strength and the $I_{MIT}/I_{M2}$ ratio in Fig. 8C. We find values in the range of 245 to 290 G.

There are considerable uncertainties associated with this technique (*40*). In addition to the usual uncertainties arising from the plasma electron density measurements and the quality of the atomic data, the accuracy of the probability (Einstein coefficient) of the MIT transition depends critically on determining the energy separation between the $^4D_{5/2}$ and $^4D_{7/2}$ levels. This is difficult simply because it is only a few milliangstroms and the EIS spectral resolution is much larger (22 mÅ). The value we use here was determined independently to be 2.29 ±0.5 cm$^{-1}$, i.e., with an uncertainty of ~20% (*58*). The largest uncertainty is, however, the EIS photometric calibration, both between the short- and long-wavelength detectors and also the evolution of the sensitivity with time. For consistency, we use the same calibration as we used for our plasma composition measurements (*54*). Considering all these factors and from a propagation of error analysis, the total uncertainty is estimated to be ~70% (*40*).

A separate issue is that as the magnetic field strength increases above ~200 G, the MIT transition affects the $^4D_{7/2}$ level population, and thus, the M2 transition becomes indirectly dependent on the magnetic field strength. In these conditions, we can use an alternative method that involves calculating the *r(E*1/184) ratio to subtract the E1 intensity contribution from the Fe X 257.262-Å line intensity. We can then model the combined M1 and MIT intensity, again using the Fe X 184.536-Å line, as a function of the magnetic field (*40*), i.e.

$$\frac{I_{MIT+M2}}{I_{184}} = \frac{I_{257} - (I_{184} \times r(E1/184))}{I_{184}} \quad (3)$$

Because this involves one less intensity ratio in the calculation, the uncertainty is ∼60%.

We show the *r(E*1/184) theoretical ratio as a function of electron density in Fig. 8B, with the values inferred from the measured electron density again overlaid. We also show the theoretical $\frac{I_{MIT+M2}}{I_{184}}$ ratio as a function of the magnetic field strength in Fig. 8D, calculated at the representative density of log (*n*/*cm*$^{-3}$) = 9.5. Using this technique, we find values in the range of 325 to 550 G. Although somewhat stronger, given the quoted uncertainties, our measurements using Eq. 3 are consistent with the field strengths we estimated using Eq. 2. Unfortunately, the range of sensitivity of this ratio to the magnetic field strength is less than the technique for weaker fields, and one of our measurements is slightly above the limit of the diagnostic capability of this method, although this would suggest even stronger magnetic fields. All of these results point toward field strengths on the order of a few hundred Gauss in all the regions.

**Acknowledgments**
We thank D. Reames for providing the Wind Sulfur data. We also thank E. Landi for providing theoretical coronal magnetic field data and advice. Hinode is a Japanese mission developed and launched by ISAS/JAXA, collaborating with NAOJ as a domestic partner and NASA and STFC (UK) as international partners. Scientific operation of the Hinode mission is conducted by the Hinode science team organized at ISAS/JAXA. This team mainly consists of scientists from institutes in the partner countries. Support for the post-launch operation is provided by JAXA and NAOJ (Japan), STFC (UK), NASA, ESA, and NSC (Norway). The HMI and AIA data are courtesy of the NASA/SDO and the AIA and HMI science teams. We would also like to thank JHelioviewer for being able to browse these data [http://jhelioviewer. org (*59*)]. We are grateful to the Wind and EPACT instrument teams for making data publicly available at the NASA Space Physics Data Facility. This research has made use of the open-source and free community-developed Python packages of SunPy v2.0.1 (*44*) and pfsspy v0.5.2 (*45,46*). This work uses GONG data from NSO, which is operated by AURA under a cooperative agreement with NSF and with additional financial support from NOAA, NASA, and USAF.

**Funding**
The work of D.H.B. was performed under contract to the Naval Research Laboratory and was funded by the NASA Hinode program. The work of S.L.Y. was carried out with support from the UKRI SWIMMR Aviation Risk Modelling (SWARM) grant.

**Author contributions**
D.H.B. designed the spectroscopic technique and analyzed EIS, Wind, and AIA observations. S.L.Y. performed the magnetic field modeling and analyzed AIA, HMI, and GOES data. Both the authors contributed to preparing the manuscript and figures.

**Competing interests**
The authors declare that they have no competing interests.

**Data and materials availability**
All data used in the paper are available publicly or from the instrument teams. The processed data needed to evaluate the conclusions of the manuscript are present in the paper. All additional data used are available from the authors.

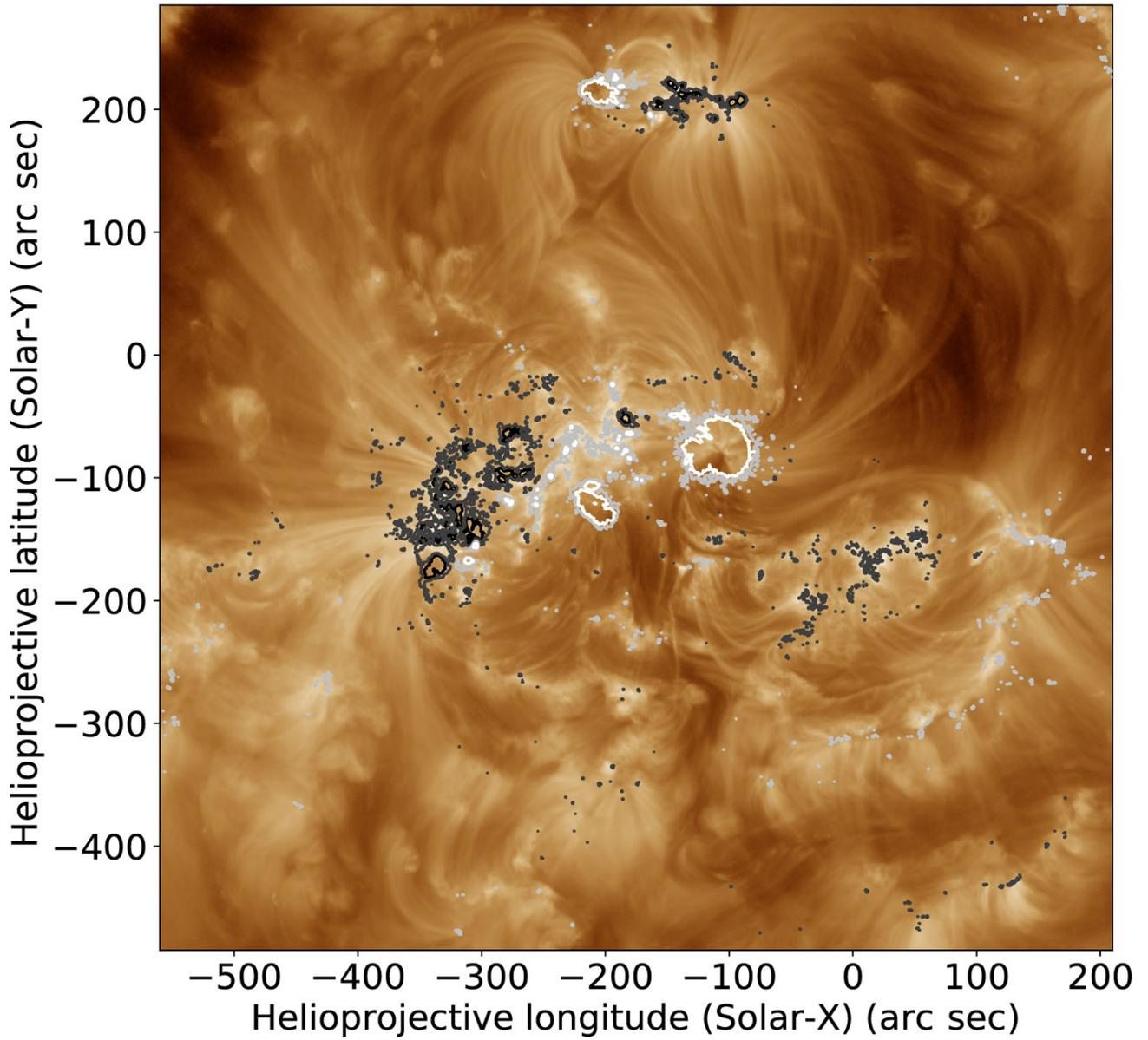

**Fig. 1. SDO/AIA 193 Å image of AR 11944.** Positive (white) and negative (black) photospheric magnetic flux contours are overlaid with a saturation of ±500 G.

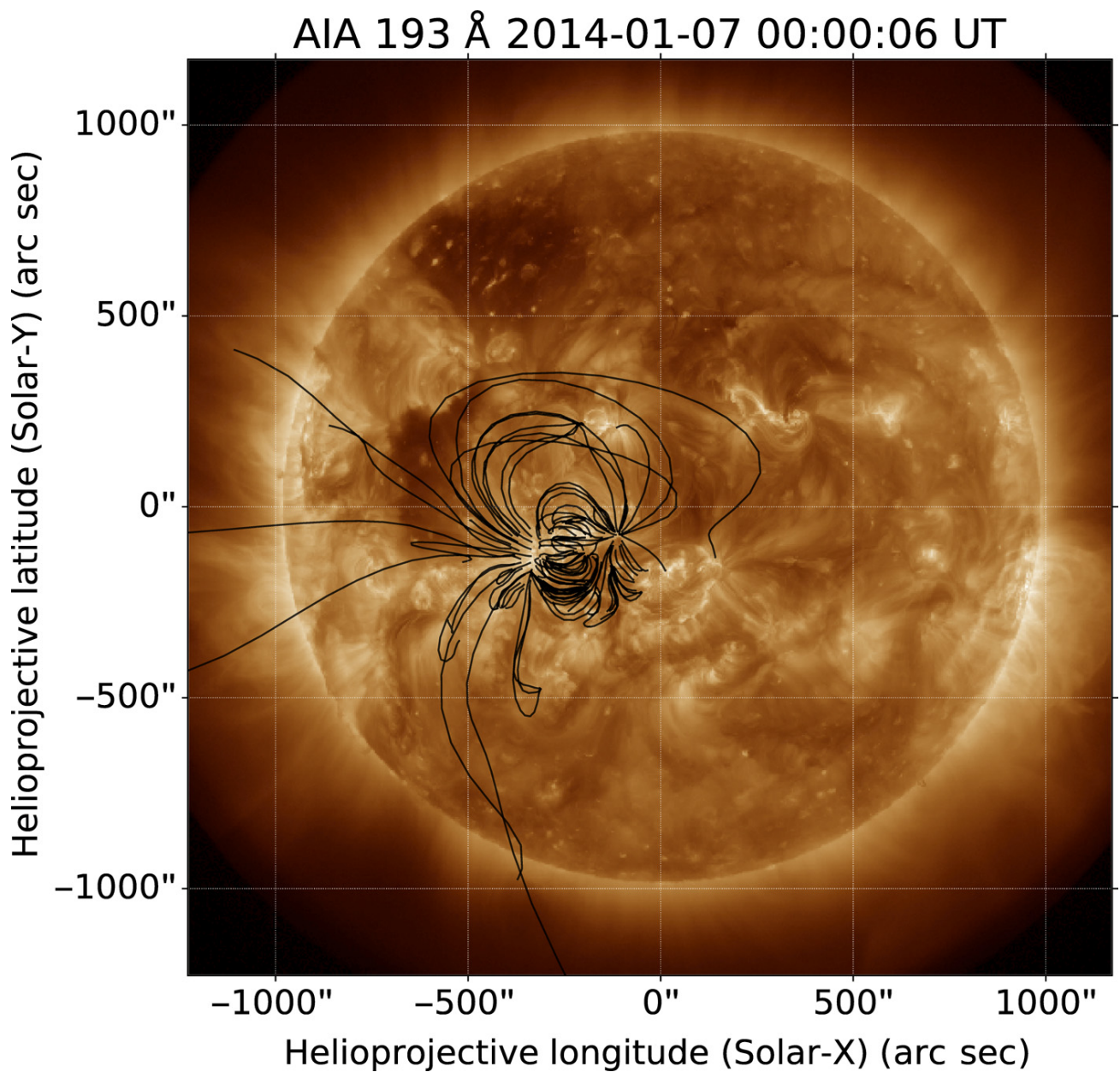

**Fig. 2. SDO/AIA 193 Å full Sun image showing AR 11944.** This image was formed at a temperature of ∼1.6 MK. We have generated a PFSS extrapolation using GONG synoptic data and overlaid some representative magnetic field lines in black.

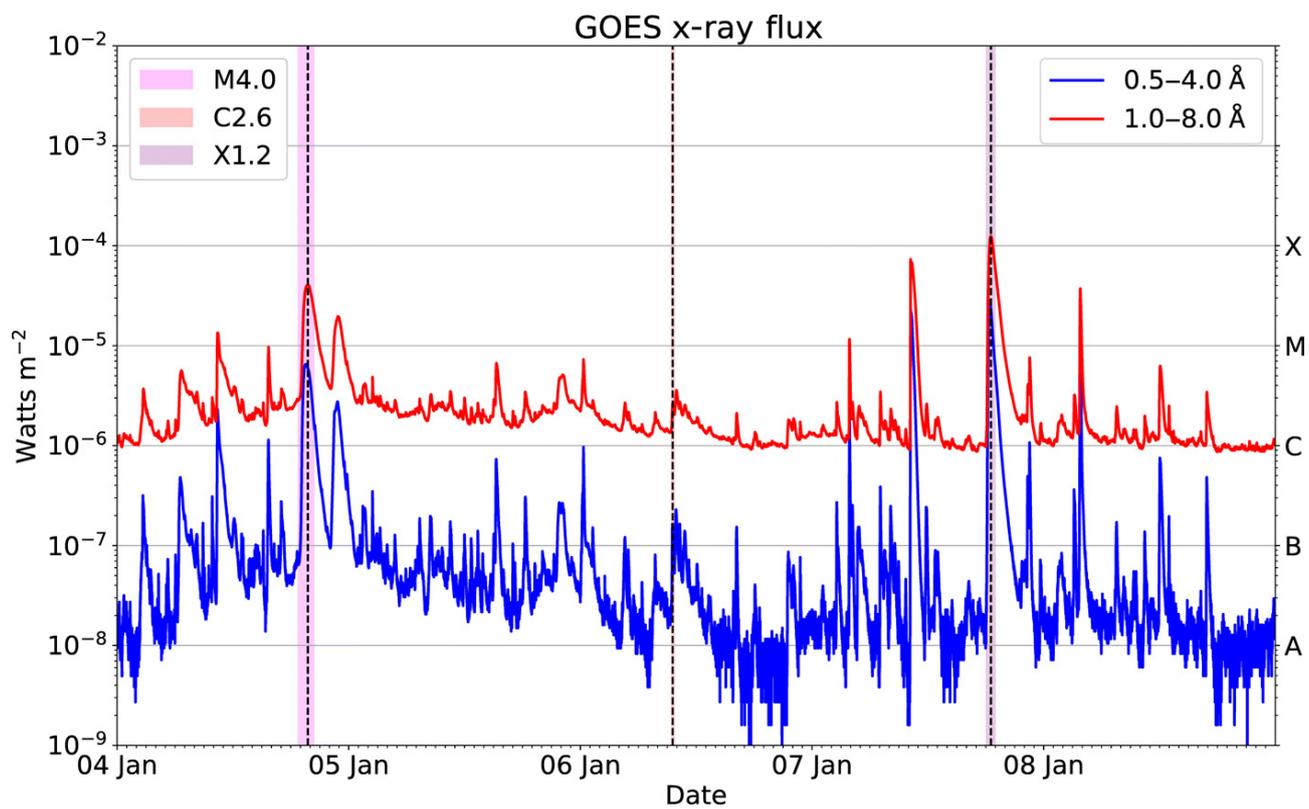

**Fig. 3. GOES x-ray flux in the 0.5- to 4.0-Å (blue) and 1.0- to 8.0-Å (red) channels.** The M4.0, C2.6, and X1.2 flares are marked by the vertical dashed lines. The shaded areas cover the start and stop times of the flares.

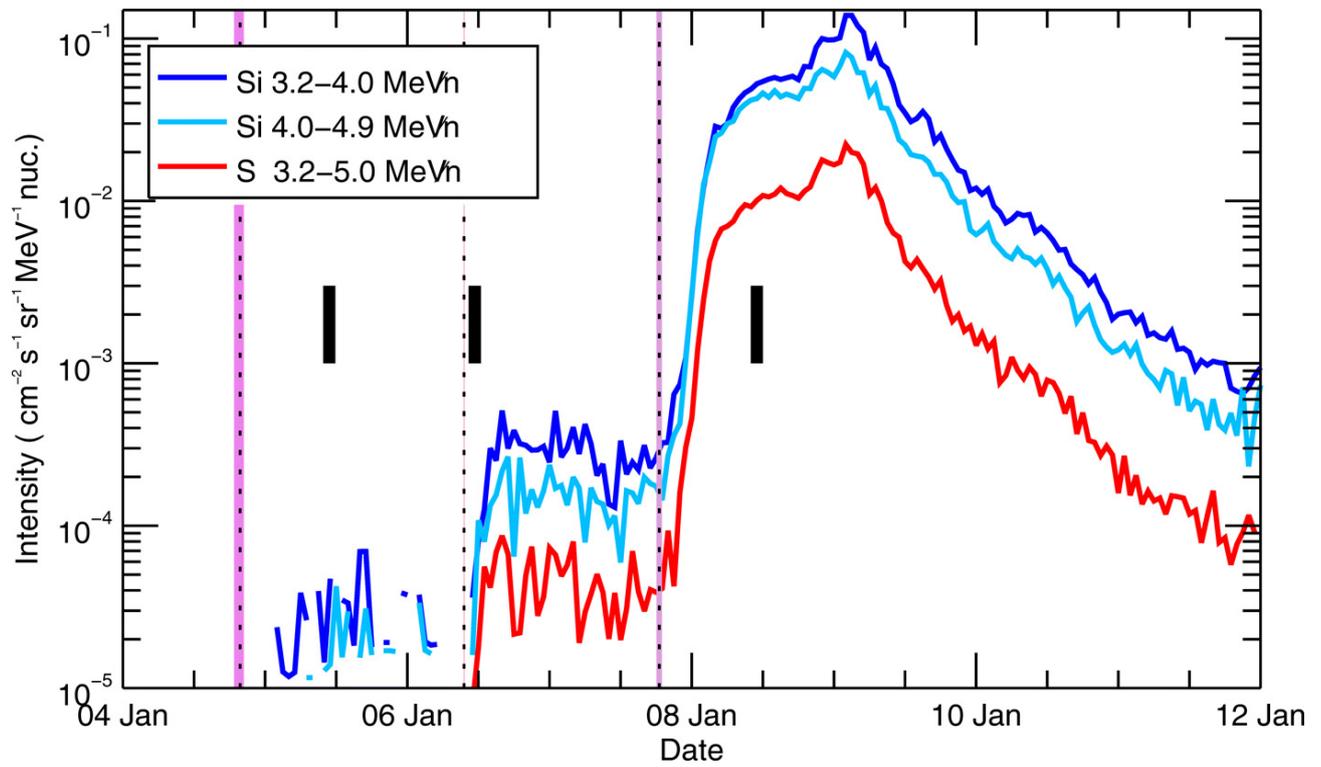

**Fig. 4. Si and S intensities measured in three energy channels by EPACT/LEMT between 4 and 12 January 2014.** The flares shown in Fig. 3 are marked by the vertical dashed lines, and the shaded areas cover the flare start and stop times. The three gradual SEP events on the 4th, 6th, and 8th appear shortly after each flare. We show the time periods of the EIS observations with the three black bars.

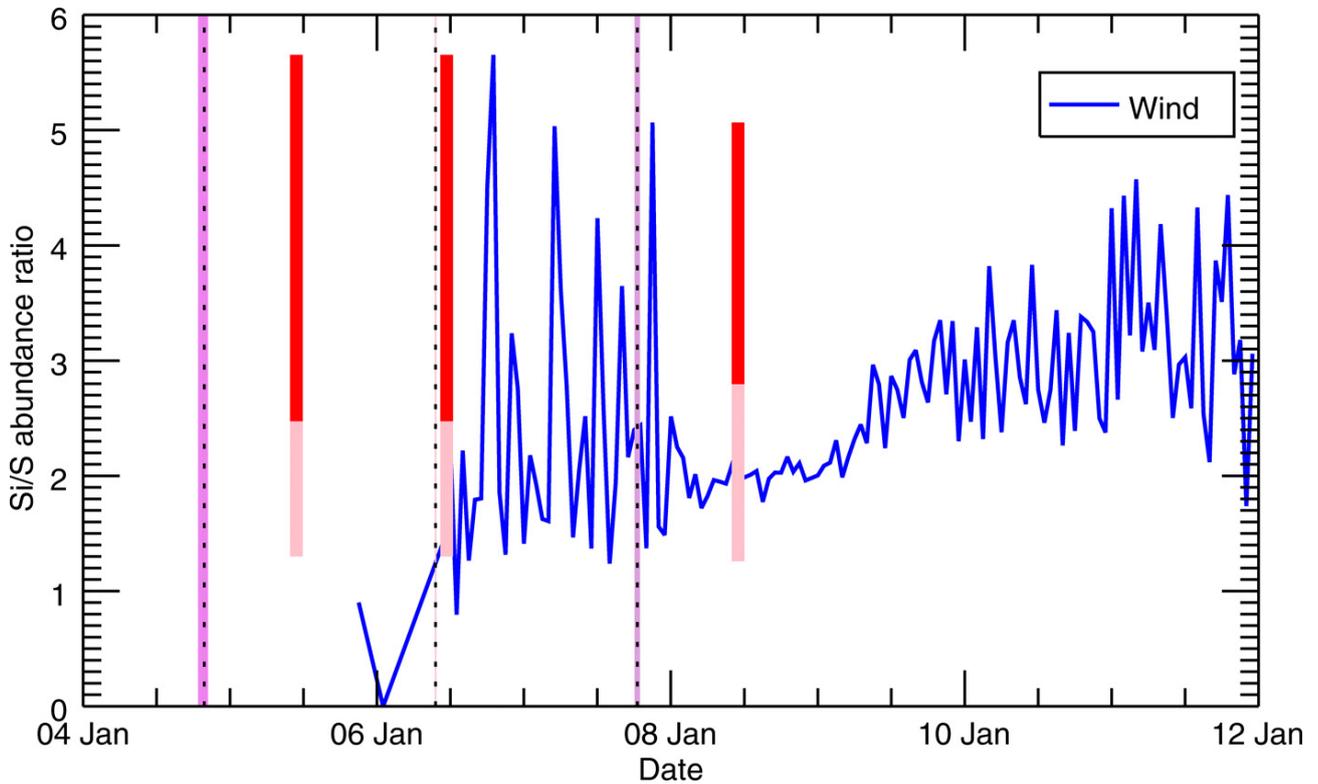

**Fig. 5.** Si/S abundance (SEP/photospheric) ratio (blue) measured by EPACT/ LEMT between 4 and 12 January 2014. The data are an average of the 3.2 to 4.0 MeV/n and 4.0 to 4.9 MeV/n channels shown in Fig. 4 normalized to the solar photospheric abundance ratio. The vertical red/ pink color bars are markers for comparison with the three EIS observations shown in Fig. 7. These EIS observations were made during the time intervals 09:48 to 11:52 UT on January 5, 10:21 to 12:25 UT on January 6, and 09:57 to 12:01 UT on January 8. The color bars are drawn on the plot to cover these time periods. The pink section of these shaded areas corresponds to the EPACT/ LEMT event average for the SEP episodes that occurred closest to each EIS observation. For example, because the S data are weak on January 5, no composition ratio measurement is possible for the event on that date. In this case, we show the EPACT/LEMT data for the second event, because it is closest to the first EIS observation. The second and third color bars, corresponding to the second and third EIS observations, respectively, show EPACT/LEMT data for the second and third events. The red sections of these shaded areas show the higher range of values up to the maximum measured by EPACT/LEMT for the same events. For cross-reference, we have overlaid the same vertical dashed lines and shaded areas that cover the flare start and stop times from Fig. 4.

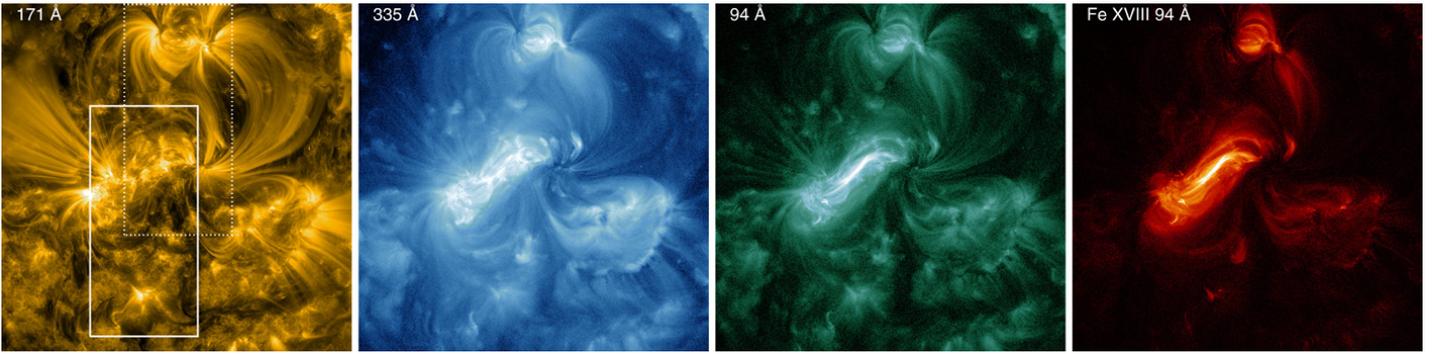

**Fig. 6. Context SDO images showing the EIS FOV.** SDO/AIA 171-, 335-, and 94-Å and Fe XVIII 94-Å images showing AR 11944 for the same FOV as Fig. 1. The 171- and 335-Å images are formed at temperatures of ∼0.9 and ∼2.8 MK, while the 94-Å images are formed at ∼7.1 MK. We isolated the Fe XVIII component from the AIA 94-Å image as described in the text. The images were taken at 11:00 UT on January 8, which is close to the time that EIS had reached the center of its scan on that date. The EIS FOV for January 8 is shown by the solid box outline. We have rotated the FOVs for the EIS scans on January 5 and 6 to the time of the 171-Å image, and we show the approximate location with the dotted box outline.

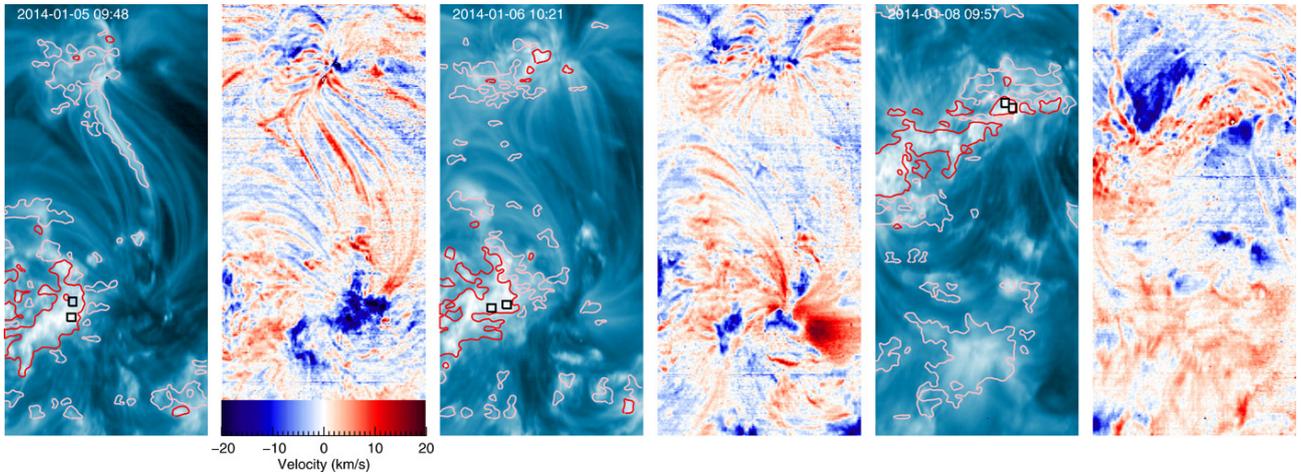

**Fig. 7. Hinode/EIS spectroscopic diagnostic maps of AR 11944.** The blue images are Fe XII 195.119-Å intensity maps. The dark blue/red images are Doppler velocity maps. Blue/red indicates plasma flowing toward/away from the observer. The color bar shows the range of velocities in the images. Contours overlaid on the intensity maps show concentrations of plasma composition within the ranges shown in Fig. 5, i.e., the pink and red contours correspond to the values of the pink and red color bars in that figure. The small black boxes show the locations where we measured the coronal magnetic field strength.

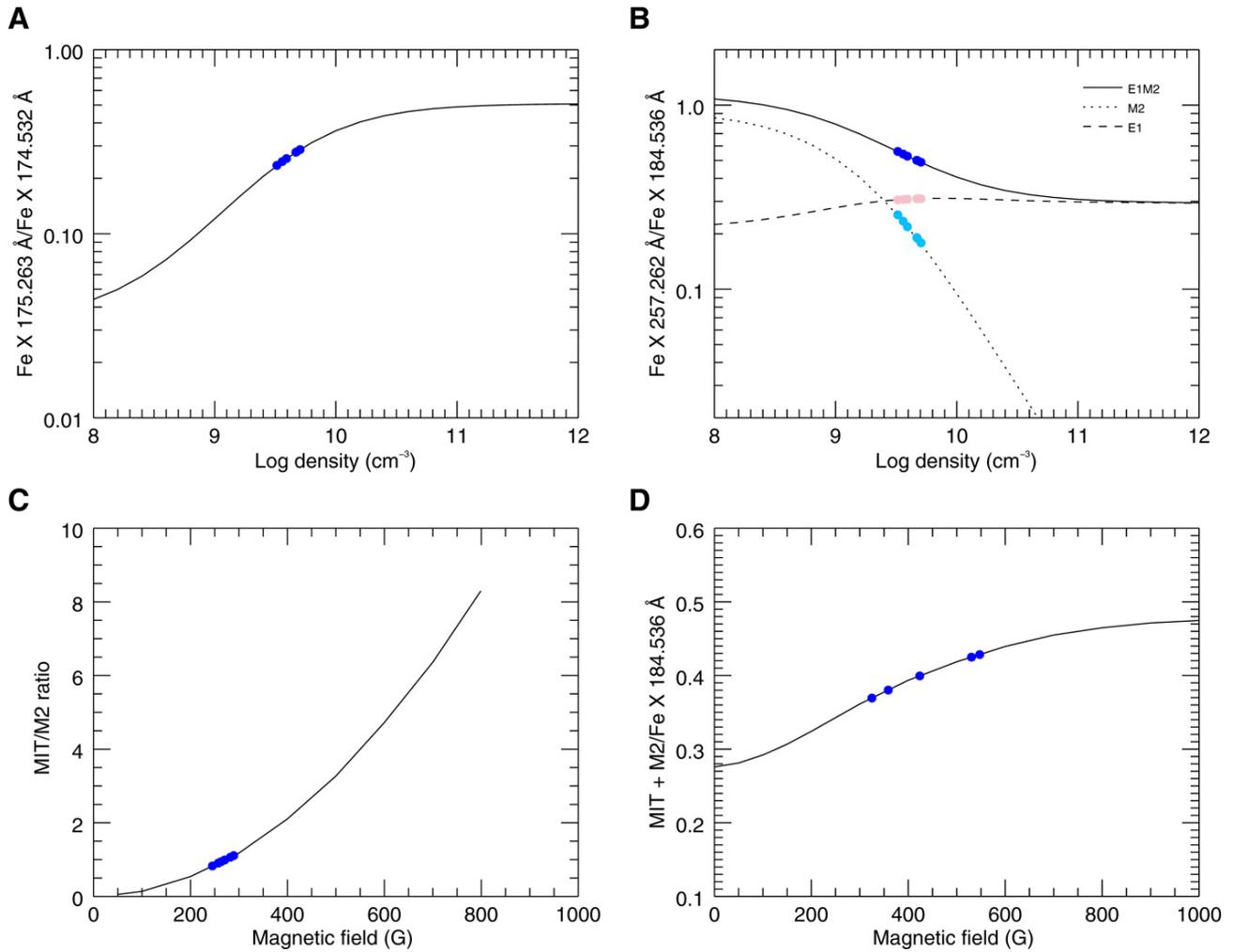

**Fig. 8 Coronal magnetic field strength measurements for the boxed regions in Fig. 7.**
(**A**) Fe X 175.263/174.532 density-sensitive ratio (solid line) with the measured values overlaid (blue dots). (**B**) The theoretical $r(E1M2/184)$ (solid), $r(M2/184)$ (dotted), and $r(E1/184)$ (dashed) ratios as a function of electron density, with the measured values overlaid (blue, sky blue, and pink dots, respectively). (**C**) MIT/M2 intensity as a function of magnetic field strength (solid line), with the values measured using **Eq. 2** overlaid as blue dots. (**D**) $\frac{I_{MIT+M2}}{I_{184}}$ ratio as a function of the magnetic field strength (solid line) for a density of log $(n/cm^{-3}) = 9.5$, measured using **Eq. 3** overlaid as blue dots.

TABLE 1
FE X LINE INTENSITIES *

| | 5-Jan-2014 | | 6-Jan-2014 | | 8-Jan-2014 | |
|---|---|---|---|---|---|---|
| λ (Å) | South | North | West | East | West | East |
| 174.532 | 5952.4±87.6 | 3416.0±67.8 | 4752.8±84.7 | 4476.8±76.4 | 5807.9±83.0 | 5390.9±90.7 |
| 175.263 | 1523.8±44.9 | 946.2±38.5 | 1117.0±40.2 | 1282.7±44.0 | 1432.6±43.4 | 1500.5±46.6 |
| 184.536 | 1425.4±9.5 | 703.2±6.3 | 1130.7±8.8 | 1042.2±8.2 | 1269.1±9.1 | 1215.9±10.2 |
| 257.262 | 1046.0±3.1 | 485.5±1.9 | 892.8±2.7 | 708.7±2.4 | 934.2±2.7 | 862.9±2.8 |

* Units of erg cm$^{-2}$ s$^{-1}$ steradian$^{-1}$.